\documentclass[conference]{IEEEtran}
\IEEEoverridecommandlockouts
\usepackage{cite}
\usepackage{amsmath,amssymb,amsfonts}
\usepackage{graphicx}
\usepackage{textcomp}
\usepackage{xcolor}
\usepackage{listings}
\usepackage{xspace}
\usepackage{enumitem}
\usepackage{url}
\usepackage{algorithm}
\usepackage{algpseudocode}

\newcommand\code[1]{\textsc{#1}}
\usepackage{caption}
\usepackage{subcaption}
\usepackage{graphicx}
\usepackage{listings}
\usepackage{xcolor}
\usepackage{xstring}
\usepackage{etoolbox}

\newcommand{\sys}{TransactionMerger\xspace}
\newcommand{\merger}{Merger\xspace}
\newcommand{\mergers}{Mergers\xspace}
\newcommand{\partitioner}{Partitioner\xspace}
\newcommand{\partitioners}{Partitioners\xspace}

\def\BibTeX{{\rm B\kern-.05em{\sc i\kern-.025em b}\kern-.08em
    T\kern-.1667em\lower.7ex\hbox{E}\kern-.125emX}}
\begin{document}

\title{Improving Database Performance by Application-side Transaction Merging\\}

\author{\IEEEauthorblockN{Xueyuan Ren}
\IEEEauthorblockA{\textit{The Ohio State University}\\
Columbus, OH, USA \\
ren.450@osu.edu}
\and
\IEEEauthorblockN{Frank Li}
\IEEEauthorblockA{\textit{The Ohio State University}\\
Columbus, OH, USA \\
li.12634@buckeyemail.osu.edu}
\and
\IEEEauthorblockN{Yang Wang}
\IEEEauthorblockA{\textit{The Ohio State University}\\
Columbus, OH, USA \\
wang.7564@osu.edu}
}

\maketitle

\begin{abstract}

This paper explores a new opportunity to improve the performance
of transaction processing at the application side by merging structurely
similar statements or transactions.
Concretely, we re-write transactions to
1) merge similar statements using specific SQL semantics; 2) eliminate redundant reads; and 3) merge contending statements across transactions by pre-computing their aggregated effect.
Following this idea, we present the design of \sys, a middleware to collect and merge transactions across
different clients. We further present a static analysis tool to identify the merging opportunity without violating isolation as well as our experience of re-writing transactions in TPC-C and Spree, a popular real-world
application. Our evaluation shows that such transaction merging can improve TPC-C throughput
by up to 2.65X and Spree throughput by 3.52X. 

\end{abstract}


\section{Introduction}

Our community has been improving the performance of database systems
in multiple dimensions, including concurrecny control~\cite{Thomson2012Calvin, Tu2013Silo}, query planning~\cite{leis2015good},
replication~\cite{Zhang2015TAPIR}, etc. However, the proposed techniques in academia
often take a long and unpredictable amount of time to get adopted into mature database systems,
probably due to the complexiy of those systems.
Similarly, commercial database systems, such as Oracle and
SQL server, may provide optimizations that do not exist in
open-source databases.

For those who urgently need better performance but cannot change their databases,
an alternative is to optimize performance at the application side,
by rewriting or restructuring its SQL transactions. Examples include
transaction chopping~\cite{Shasha1995Chopping}, re-ordering statements~\cite{Xie2015HAV}, transaction sharing~\cite{rehrmann2020sharing}, etc.
Compared to database-side optimizations, application-side optimizations may require additional effort from
the application developer, but they may be deployed in a predictable amount of time.
Furthermore, application-side optimizations can leverage application-specific
knowledge that is often hard to obtain at the database side. Prior studies
have shown that such application-side optimizations are quite common in real-world applications~\cite{dbcc-cidr-2023,tang2022adhoc}.

This paper explores a new application-side optimization opportunity by merging 
structurely similar statements or transactions.
Compared to the naive client-side batching approach~\cite{oracleStatementJava}, which
executes a few insert, update, or delete statements together, our
approach is more intrusive, requiring changes to the application
code, but gains more improvement. To differentiate our approach
from the naive one, we call our approach “transaction merging” in
this paper. In particular, this paper explores two opportunities:

\vspace{.05in}
\noindent
\textbf{Intra-transaction merging.} A transaction can have multiple 
\code{insert}, \code{update}, or \code{delete} statements to change the states of multiple
rows, or multiple \code{select} statements to check multiple conditions. Since these
statements often do not have dependencies among each other, they create a 
natural opportunity for merging. We observe that SQL has already
provided semantics to support merging: We can concatenate the conditions
of multiple \code{select} or \code{delete} statements to execute their tasks together;
\code{insert} supports insertion of multiple rows; and we can use \code{update case when}
statement to update multiple rows with different conditions.

Our micro-benchmark evaluation shows that such intra-transaction merging can improve
transaction processing throughput by up to 6.0X on MySQL and 4.9X on PostgreSQL.

\vspace{.05in}
\noindent
\textbf{Inter-transaction merging.} Database applications often expose a number of
APIs to the end users, each executing one or a few transactions. Examples include
the \code{new-order} and \code{payment} transactions in TPC-C~\cite{tpccurl}. As a result,
if multiple users are accessing the application concurrently, they may issue transactions 
from the same API, which are structurely similar but have different argument values.
Such transactions provide another opportunity for merging. For example, we can
implement a transaction to process multiple \code{new-order} or \code{payment} requests from
multiple customers, using the SQL semantics discussed above.
To realize this idea, we have built \sys, a middleware that can accumulate similar transactions from clients,
merge them into one, execute the merged transaction, and dispatch results to the clients.

Inter-transaction merging brings at least two more optimization opportuities compared to intra-transaction
merging. First, many clients' transactions read the same data.
For example, a \code{new-order} transaction in TPC-C needs to read the tax rate of the local
warehouse, and thus customers of the same warehouse all read the same value. Therefore,
if we merge \code{new-order} from the same warehouse, the merged transaction only needs to read tax rate once, eliminating the redundant reads.
Second, inter-transaction merging can turn contention, which is typically
a bottleneck in transaction processing, into an optimization opportunity.
For example, if two transactions both want
to deduct a number by one, the merged transaction can simply deduct the number
by two---it not only eliminates the contention, but also reduces the number of operations
the database needs to execute.

However, inter-transaction may break the boundary of transactions, raising many questions~\cite{rehrmann2023merging}. For example, in a locking-based
concurrency control mechanism, if two transactions both want to update the same row, one
will be delayed---this is necessary to achieve isolation levels like SERIALIZABLE, etc.
If we simply merge two updates from two transactions, they may generate results that would
not occur under SERIALIZABLE isolation level. To address this problem,
we have built a static analysis tool to help developers identify the
opportunities of inter-transaction merging without breaking isolation.

In this paper, we make the following contributions:

\begin{itemize}[leftmargin=*]

\item We design and implement \sys, a middleware to collect and merge transactions from multiple
clients. Its design follows two principles. First, \sys should be stateless
to simplify fault tolerance, load balancing, and scaling. 
Fortunately, we find the abort support for database transactions greatly simplifies
this task, as \sys can always abort a transaction that fails in the middle, instead
of trying to persist its intermediate states to resume its execution.
Second, \sys should support online
re-configuration, as the optimal setting (e.g., the number of transactions to merge, which
transactions to merge, etc) often depends on the workload and thus may change over
time. To avoid the expensive synchronous configuration update, our design ensures
that the correctness of \sys will not be violated by the temporal inconsistency between
different components.

\item 
We design a static analysis tool to help developers identify inter-transaction
merging opportunities without breaking isolation. Upon that,
we present our experience of implementing merged transactions for TPC-C, a standard database benchmark,
and Spree~\cite{spreecommerceWhatSpree}, a popular online shopping framework. 
Our experience shows that, first, though such effort requires
non-trivial effort,  a developer can
apply such effort incrementally, starting from a subset of performance-critical transactions
or statements. Second, beyond the opportunities suggested by our
static analysis method, a developer may leverage
application information for further optimization. This is 
an advantage of application-side optimization, which can
hardly be implemented at the database side.

\item Our evaluation shows that transaction merging can improve the throughput of TPC-C
by up to 2.65X and that of Spree by up to 3.52X, without any changes to the database backend.
Therefore, we believe \sys provides a valuable alternative for improving database
performance, especially when the developer has no control of the database backend.

\end{itemize}

\section{Opportunities}
\label{sec:opportunities}

\subsection{Intra-transaction Merging}
\label{sec:intra}

It is not uncommon for a transaction to include multiple independent \code{insert}, \code{update}, \code{delete}, or \code{select}
statements, which creates a natural opportunity for batching (see Section~\ref{sec:app} for some examples).

Actually, today's client-side SQL connector, such as JDBC, already provides \code{addBatch} and \code{executeBatch} functions to 
allow batching. When the application calls \code{addBatch}, the connector will buffer
the following \code{insert}, \code{update}, and \code{delete} statements (note that JDBC batching does not support \code{select}). 
Then when the application calls \code{executeBatch}, the connector
will send all buffered statements to the database together. As shown later, this approach
can bring performance improvement on some database implementations, but our approach can achieve more.

Our approach re-writes those transactions into one using specific SQL semantics discussed below. 
We call the statements before re-write the \emph{original statements} and the ones after re-write the
\emph{merged statements}.
To be specific,

\vspace{.05in}
\noindent
\textbf{\code{select} and \code{delete}}. 
For these statements, we can merge them in two ways. First, we can 
merge their \code{where} clause with \code{or} keyword. Second, if their
where clauses only include the equal relation, then we can use the \code{in}
keyword to merge the values their equal relation compares to.
Our evaluation shows that, when using
\code{in} is possible, it performs better than using \code{or}, so we use \code{in}
when possible.

For \code{select},
we further need to scan the result set of the merged statement to see which rows fit with which \code{where} clause and dispatch these
rows correspondingly. Since the result set is typically not large, the additional overhead is low.
In more complex scenarios where results cannot be easily separated (e.g., \code{select sum}),
we find it is usually possible to use the \code{group by} keyword to automatically separate the results.

\vspace{.05in}
\noindent
\textbf{\code{update}}. We use the \code{update case when} statement to update multiple rows. 
The \code{case-when-then-else} block is like an \code{if-then-else} that evaluates a condition and returns the corresponding value. 
We put all conditions in the \code{where} clause into several \code{when} blocks, where the return value in each \code{then} block is exactly the value updated for each condition. 
We set the value in the \code{else} block to be the original value in case no condition is met.

\vspace{.05in}
\noindent
\textbf{\code{insert}}. \code{insert} statements with \code{values} keyword can insert multiple rows
separated by commas.
If there are multiple \code{insert} statements that can be merged, 
we concatenate them into one statement with all value lists appended.

\begin{figure*}[t]
    \centering
    \begin{subfigure}[b]{0.24\textwidth}
        \centering
        \includegraphics[width=\textwidth]{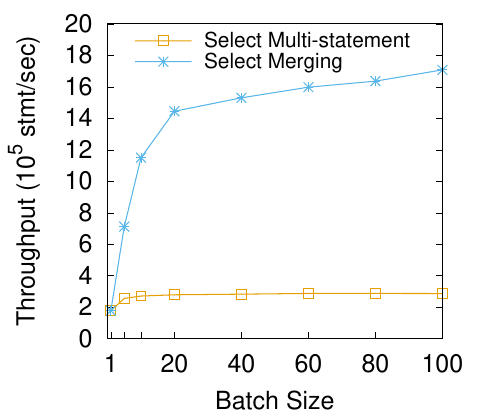}
        \caption{\code{select}-MySQL}
        \label{fig:micro-select}
    \end{subfigure}\hfill
    \begin{subfigure}[b]{0.24\textwidth}
        \centering
        \includegraphics[width=\textwidth]{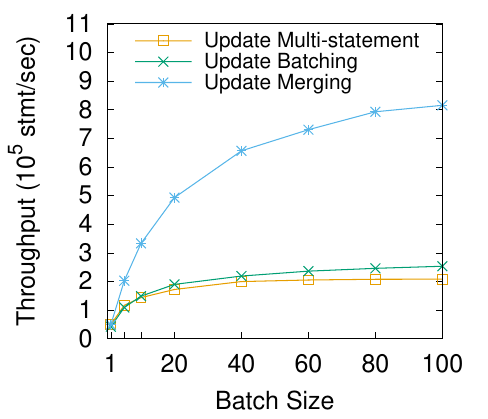}
        \caption{\code{update}-MySQL}
        \label{fig:micro-update}
    \end{subfigure}\hfill
    \begin{subfigure}[b]{0.24\textwidth}
        \centering
        \includegraphics[width=\textwidth]{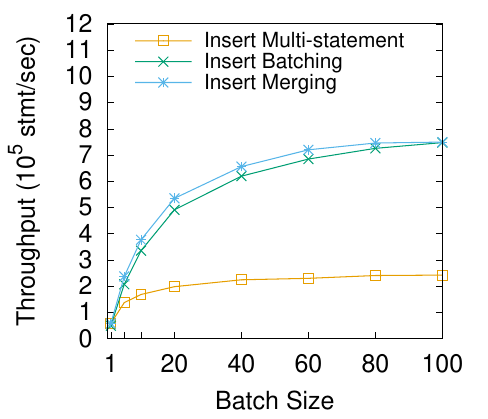}
        \caption{\code{insert}-MySQL}
        \label{fig:micro-insert}
    \end{subfigure}\hfill
    \begin{subfigure}[b]{0.24\textwidth}
        \centering
        \includegraphics[width=\textwidth]{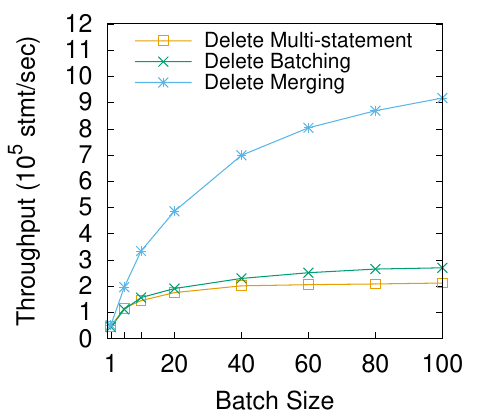}
        \caption{\code{delete}-MySQL}
        \label{fig:micro-delete}
    \end{subfigure}
     \caption{Throughput of the micro-benchmark on MySQL.}
    \label{fig:micro-mysql}
\end{figure*}

To measure the potential benefits of intra-transaction merging, we perform a microbenchmark evaluation.
It creates a table with one million rows for \code{select}/\code{update} and 15 million rows for \code{delete} 
(we need larger number of rows for \code{delete} since \code{delete} reduces the number of rows during the experiments), and each row has two columns: id and value. In the original version,
each transaction executes a batch size of statements of the same type (i.e., \code{select}, \code{insert}, \code{update}, or \code{delete}),
each statement operating on a random single row using the primary key (id). We use the following for each statement:
1. \code{select value from table where id = ?};
2. \code{insert into table (id, value) values (?, ?)};
3. \code{update table set value = ? where id = ?};
4. \code{delete from table where id = ?}.
We measure the throughput of the original version in terms of the number of statements per second.
For \code{insert}, \code{update}, or \code{delete}, we also measure the throughput with JDBC batching.

Then our approach merges all statements in each transaction into a single merged statement, using
the method described above. 
 For a fair comparison with the original version, 
we report the throughput of the merged version as transactions per second times the number of statements 
to be merged per transaction. The hardware and software configurations of these experiments can be found in Section~\ref{sec:eval}.

Figure~\ref{fig:micro-mysql} compares the throughput of original and merged versions on MySQL. As shown in this figure,
MySQL can merge multiple \code{insert} statements into one, like our merging, but does not do much optimization for \code{update} and \code{delete}. Our merged version outperforms
MySQL on \code{select}, \code{update}, and \code{delete}, bringing up to 6.0X improvement.
We have performed a similar measurement on PostgreSQL and found our version
outperforms PostgreSQL
on \code{select} and \code{update} workloads, achieving up to 4.9X and 1.4X speed up respectively.

\subsection{Inter-transaction Merging}
\label{sec:inter}

A middleware between the database clients and the database servers
can potentially merge transactions of the same type from different clients.
This approach brings both benefits and challenges.

On the benefit side, first, inter-transaction merging brings more opportunities
for merging. A single transaction only has a limited number of statements for
merging, but when considering transactions of the same type from different
clients, the number of statements that can be merged increases significantly.

Furthermore, when different transactions contend, which means they access
the same rows, the contending statements can be merged more effectively
than non-contending statements. For example, if two transactions from two
customers both want to purchase the same item, which means they both
need to decrement the item count by one, the merged version can simply
decrement the item count by two. In this way, two updates can be merged into
one---this is even better than merged non-contending transaction, which still
needs to update two rows.
With this approach, we can turn contending transactions, which are typical
challenges for database performance, into optimization opportunities that are
more beneficial than non-contending transactions.

Finally, some transactions may perform the same checks (e.g., item is available) or
 read the same values (e.g., date or tax information). When merged, the merged
version only needs to execute those operations once.

The potential improvement of these scenarios can be inferred from intra-transaction merging results (Figure~\ref{fig:micro-mysql}). For example,
assuming each original transaction updates a single row, its throughput will be equal to
the (batch size=1) point of the original or batched line in the figure; assuming they are not contending and we
can merge X of them, the throughput of the merged version will be equal to
the (batch size=X) point of the merged line in the figure; assuming they are contending and
we can merge X of them, the throughput will improve by X times.

Implementing inter-transaction merging, however, is more challenging than implementing
intra-transaction merging, due to several reasons. First, transaction is the basic unit for
atomicity and isolation in a database system. Unlike intra-transaction merging, inter-transaction
merging is breaking the boundary between transactions, which causes problems to isolation.

To address these problems, we have built a static analysis tool to help developers find merging
opportunities without breaking isolation (Section~\ref{sec:static}). Furthermore,
our experience is that re-writing code at the application level
leveraging application-level knowledge and constraints can gain more improvement.
For example, consider a transaction that first increments an ID by one
and then uses the new ID to create an order so that different orders have
different IDs.  The ``increment by one''
statement is a typical contention point as multiple transactions may
try to update the same value. However, we cannot simply
merge multiple ``increment by one'' into a single ``increment
by n'', since then multiple transactions will use the same ID to
create orders. The problem is caused by the fact that merging is breaking the isolation
boundary of transactions: In the original transaction, one transaction
can only see ``increment by one'' from itself, not from othe transactions;
in the merged version, the transaction can see the increments from all
other transactions.
On the other hand, at the application level,
re-writing is pretty straightforward. A merged transaction
can read the old ID value into a local variable (old\_id),
increment it by n in the database, and then use old\_id+1,
old\_id+2, etc, to create each order.
However, determining how to do this automatically probably
would require complex analysis.

The second challenge is how to determine which transactions to merge. Our observation is that it is
better to merge contending transactions, because doing so will have more improvement as discussed above
as well as allowing different transactions to execute in parallel. In theory, leveraging this idea is not easy, as different
transactions may contend in different ways and at multiple locations, and such information
may not be available before a transaction is executed. The applications we use in this work, however,
include easy-to-identify merge points, such as the warehouse district in TPC-C. This is probably
because, when partitioning data in a distributed database, the database designer already tries to minimize 
cross-partition transactions in the first place. As a result, we can naturally try to
merge transactions to the same partition. However, when necessary, it is possible to
apply more sophisticated analysis to determine which transactions to merge (see Section~\ref{sec:related}).

\section{Design and Implementation}
\label{sec:design}

 \begin{figure*}[t]
         \centering
            \includegraphics[width=0.7\textwidth]{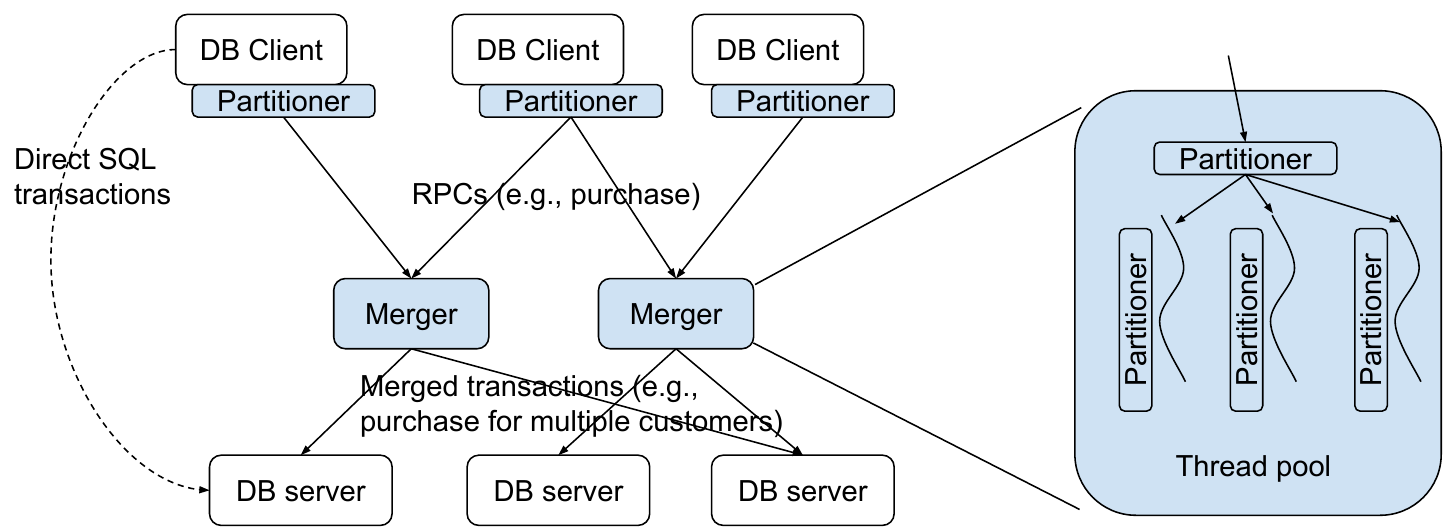}
             \caption{Design of  \sys}
             \label{fig:design}
\end{figure*}

Figure~\ref{fig:design} presents the design of \sys. As shown in this figure, \sys is a typical middleware
 between the database clients (e.g., web servers) and database servers.
The developer needs to submit the code of the merged version to \sys.
The database client will send RPCs to the \mergers in the middleware (e.g., \code{purchase(item=book1,item\_count=3)}). 
Each \merger will collect a number of RPCs, merge them, and execute the merged version submitted by the developer,
which involves sending merged SQLs to the database server and dispatching the result back to the clients.
Internally every \merger
can have multiple threads to maximize performance.

As discussed in Section~\ref{sec:inter}, it's better to merge transactions with certain properties.
To maximize the chance of merging, \sys incorporates \partitioners at different locations
to ensure transactions that may be merged can go to the same thread. For example, in TPC-C,
it's a natural idea to merge transactions to the same warehouse district. Then the client-side \partitioner
may apply policies like transactions to warehouses 1-10 go to \merger 1, transactions to warehouses 11-20 go to \merger 2, and so on.
Similarly, \merger 1's \partitioner may apply policies like transactions to warehouses 1-2 go to thread 1,  transactions to warehouses 3-4 go to thread 2,  and so on. Finally, each thread's \partitioner partitions transactions based on their warehouse districts.

Alternatively, a client can also send SQL transactions directly to the server.
This allows a developer to incrementally merge a subset of transactions in \sys.

\sys's design follow two principles to achieve fault tolerance, scalability, and performance.

\vspace{.05in}
\noindent
\textbf{Keep \mergers stateless.} A service is stateless if it does not need to keep any
persistent data for it to function properly. A stateless service has multiple benefits for simplicity of design:
If a service instance fails, we just need to create a new blank instance, which means we do not need
to pay the cost of replication, etc; a client can connect to any service instance, which simplifies load balancing,
as we can freely migrate clients across different service instances; 
if the service experiences an increase of traffic, the service can add blank instances for scaling.

To keep \mergers stateless, \sys still relies on the database backend to achieve atomicity, isolation, durability, etc. 
As a result,  a \merger does not need to maintain data for mechanisms like redo logs, concurrency control, etc,
which significantly simplifies its design. If a \merger fails for
any reason, instead of trying to resume its unfinished transactions, which would require the \merger to replicate its intermediate states,
\sys can always abort these unfinished transactions and ask the clients to retry.
Of course aborting and retrying incur additional overhead, but since \merger
failures should be rare, such overhead should be negligible compared to the cost of replicating intermediate states.

The most challenging case is when a \merger finishes a merged transaction successfully, but fails before dispatching the results
to the clients. However, we observe this case is no different from the case that a client sends a direct transaction to
the database server, the server finishes it, and the client fails to receive the result either due to network problems or due to
client failure. In other words, the client application should already have logics to handle this case, no matter
whether it uses \sys or not. In practice, we observe a common solution is to maintain a flag or timestamp for each
row, marking whether the row has been processed or which transaction has processed it.

\vspace{.05in}
\noindent
\textbf{Support online re-configuration.} 
\sys may need to be re-configured for multiple reasons. First, as discussed above, if we add \mergers or remove \mergers, either
for fault tolerance or for scalability reasons, then the policies of different \partitioners need to be updated.
Second, the performance of transaction merging depends
on parameters like the number of threads, the number of transactions to merge, and the merging policy (i.e., what
transactions to merge). The optimal setting
may depend on the traffic pattern. As traffic pattern can change over time, it would be better if \sys can change such
settings at run time. 

However, naively changing such settings may incur subtle synchronization problems. Using the prior TPC-C example,
suppose \merger 1 initially was assigned warehouses 1-10, and then due to load balancing, we need to change the
policy to send transactions of warehouses 1-20 to \merger 1. In this case, both the client-side \partitioner and the
\merger-side \partitioner need to update their policies. If not carefully handled, the inconsistency between different
\partitioners may cause trouble. One solution is to implement atomic update of multiple \partitioners, but this is complicated
and can introduce a significant overhead.

To support online re-configuration, our design principle is that the correctness of the system should not depend on the
consistency across different \partitioners. In other words, every \partitioner should be able to handle every transaction,
instead of assuming that it only needs to handle a subset due to the policy of the higher-level \partitioner.
Using the prior TPC-C example, following
this principle, thread-side \partitioner should be able to partition transactions from any warehouse; \merger-side \partitioner
should have a backup plan for transactions that it is not supposed to handle (e.g., transactions from warehouses larger than 10
should go to thread X). Using this strategy, the correctness of the system will not depend on the partitioning policy. As
a result, different partitioners can be updated asynchronously; the temporary inconsistency between \partitioners may 
hurt performance, but not correctness.

With the support of online re-configuration, we have implemented a Bayesian Optimization (BO) method to
tune the number of threads and the number of transactions to merge at run time. Such method is widely
used in system configuration tuning~\cite{alipourfard2017cherrypick, behzadOptimizingPerformanceHPC2018, agarwalActiveLearningbasedAutomatic2019, rajeshTunIOAIpoweredFramework2024}. 
To be specific, we employ a Gaussian Process Regressor (GPR) to model the relationship between configurations and their corresponding throughput. 
Starting with an initial set of random configurations, we measure the throughput of the system and train the GPR model.
BO then iteratively selects new optimal configurations, retraining the GPR with new configurations.
This loop continues until throughput improvements stabilize or a predefined iteration limit is reached, 
enabling \sys to autonomously converge on high-performance configurations efficiently.

\vspace{.05in}
\noindent
\textbf{Implementation.}
We implement \merger using the gRPC framework~\cite{grpcGRPC}, exposing every transaction type
as an RPC method. A \merger also includes a number of worker threads to merge transactions,
each associated with an incoming queue. When a gRPC thread receives an RPC request, it
will determine which worker thread to send this request to using the \merger-side \partitioner,
and then enqueue the request to the corresponding queue of that worker thread.
Each worker thread will repeatedly pull requests from its own queue, partition them using
the thread-side \partitioner, merge transactions in the same partition, and send the merged
version to the database. 
The number of worker threads and the number of transactions to merge can be configured at run time
through another RPC method that allows tuning of these parameters from client side.
A thread stops merging either when it has already merged sufficient number of transactions or
when a timeout occurs.

We implement the BO method in Python, by leveraging scikit-learn's GaussianProcessRegressor 
with an RBF + WhiteKernel kernel~\cite{GPR}, where RBF captures non-linear relationships 
and WhiteKernel accounts for noise. BO via scikit-optimize's gp\_minimize~\cite{BO} explores 
a discrete parameter space, guided by the GPR's predictions. We also use a gRPC server to
integrate the re-configuration part with the \sys.

\section{Case Study}
\label{sec:app}

\lstset{
    language=SQL,
    morekeywords={FOR, IF, ELSE}, % add FOR as a keyword
    % basicstyle=\ttfamily\normalsize,
    keywordstyle=\color{blue},
    commentstyle=\color{gray},
    deletekeywords={date, DATE, number, NUMBER},    % Remove from main keywords
    deletendkeywords={DATE, NUMBER},        % Remove from ndkeywords (data types)
    % stringstyle=\color{red},
    frame=tb, % Top and bottom rule
    breaklines=true,
    showstringspaces=false,
    basicstyle=\ttfamily\scriptsize,
    showlines=true,              % Force empty lines to be printed
    emptylines=3
}

In this section, we first present our methodology to find merging opportunties and then present our
experience of implementing intra and inter-transaction merging in two applications: TPC-C and Spree. 
TPC-C is a widely used benchmark for evaluating online transaction processing (OLTP) systems, and Spree is an open-source online shopping application.
We first describe the original transactions in two applications and then show how we re-write them to implement intra and inter-transaction merging.

\subsection{Static Analysis to Find Merging opportunities}
\label{sec:static}

The key idea of this work is to merge structurely similar statements from multiple instances of the
same transaction. However, this may cause a violation of isolation between these instances.
Intuitively, given two instances of the same transaction with statements $\{S^1_1, S^1_2,...,S^1_n\}$ 
and $\{S^2_1, S^2_2,...,S^2_n\}$, we want to know that whether a serial execution of their statements
$\{S^1_1, S^1_2,...S^1_n,S^2_1, S^2_2,...S^2_n\}$ is equivalent to a re-ordered exectuion
$\{S^1_1, S^2_1,S^1_2, S^2_2,...,S^1_n,S^2_n\}$: If yes, we can apply the re-ordered execution
and create a merged execution $\{S^1_1+S^2_1,S^1_2+S^2_2,...,S^1_n+S^2_n\}$ (``+'' indicates
merging). However, the answer is not always yes. Suppose $S^1_2$ updates a value read by $S^2_1$,
then the above re-ordering will cause $S^2_1$ to get a value different from that in the serial execution, violating
isolation.

To systematically identify merging opportunities without violating isolation, our key idea is to divide
a transaction's statements into multiple groups, so that conflicts (i.e., access the same row and at least
one access is a write) only exist within a group but do not exist across groups. Then we can re-order
groups as discussed above without violating isolation. In the above example, suppose $S_1$ and $S_2$
have a conflit, but other statements do not have any conflicts, we will divide the transaction into
$\{G_1, S_3, ... ,S_n\}$ in which $G_1$ includes $S_1$ and $S_2$. Then for two instances of the
same transaction, we will re-order its execution into $\{G^1_1, G^2_1,S^1_3, S^2_3, ...,  S^1_n, S^2_n\}$.
The correctness of this re-ordering can be proved easily using the theory of serialization graph: In the dependency
graph corresponding to the re-ordered execution, there can only be conflict edges from $T_1$ to $T_2$,
because a group can only conflict with itself and we always place $T_1$'s group before $T_2$'s group of
the same position.

To identify such groups, our algorithm is to start by creating a group for each statement and then
merge groups: If two groups have a conflict or have an overlap, we merge these two groups and any
groups between them. We repeat this till no further merging is applicable.

In the re-ordered execution, if a group contains a single statement, we can merge corresponding groups with the techniques
discussed in this paper. Otherwise, we either need to execute corresponding groups sequentially or have to re-write
application logic to make merging possible. In the above example, this step creates an execution
$\{G^1_1, G^2_1,S^1_3+S^2_3, ...,  S^1_n+ S^2_n\}$.

Our current implementation relies on the developer to provide statement information, including which columns
the statement reads or writes.
Then our algorithm detects conflict with the criteria that two statements
access the same column with at least one write. There are a few exceptions:
First, if one statement inserts a new row and another one queries an existing row,
we do not consider them as a conflict. Second, if a developer is sure that two transactions
use different primary keys as the condition for some statements, those statements
will not be considered for grouping, even if they touch the same column, since they
will not create any dependencies across transactions.

\subsection{TPC-C}
TPC-C models a wholesale supplier environment and defines five transaction types, each simulating real-world business operations by specific SQL statements.
We implement our merged transactions based on an open-source tool Benchbase~\cite{benchbase} that implements TPC-C. 
Below transactions are part of Benchbase TPC-C workload.

\vspace{.05in}
\noindent
\textbf{\code{new-order}} transaction simulates a customer placing a new order into the system.
It involves retrieving customer data, fetching and updating fields related to the district's order numbering,
inserting rows for the new order and order table, and then, for each item in the order, fetching item and stock details, 
updating stock information, and inserting corresponding order line records.

\begin{figure*}[!t]
\begin{subfigure}[t]{0.48\textwidth}
\begin{lstlisting}
Input: warehouse id (w_id), district id (d_id), 
    customer id (c_id), array of item id (item_ids), 
    array of supplier w_id (supplier_w_ids), 
    array of quantity (quantities)
-- fetch customer data 
1 SELECT c_discount,... FROM customer 
    WHERE c_w_id = ? AND c_d_id = ? AND c_id = ?;
-- fetch warehouse data
2 SELECT w_tax FROM warehouse WHERE w_id = ?;
-- fetch district d_next_o_id (@next_id) for update
3 SELECT d_next_o_id AS next_id, d_tax FROM district 
    WHERE d_w_id = ? AND d_id = ? FOR UPDATE;
-- increment d_next_o_id by one
4 UPDATE district SET d_next_o_id = d_next_o_id + 1 
    WHERE d_w_id = ? AND d_id = ?;
-- insert order and new_order record
5 INSERT INTO oorder (o_id, o_w_id, o_d_id,...) 
    VALUES (@next_id,?,?,...);
6 INSERT INTO new_order (no_o_id, no_w_id, no_d_id) 
    VALUES (@next_id,?,?);
-- execute for each order_line item
for each order_line item:
  -- fetch item data
  7 SELECT i_price, i_name, i_data FROM item 
      WHERE i_id = ?;
  -- fetch stock data
  8 SELECT s_quantity,... FROM stock 
      WHERE s_w_id = ? AND s_i_id = ? FOR UPDATE;
  -- insert order_line record
  9 INSERT INTO order_line 
    (ol_w_id, ol_d_id, ol_o_id,...) 
    VALUES (?,?,?,...);
  -- update stock
  10 UPDATE stock SET 
       s_quantity = ?, 
       s_ytd = s_ytd + ?, 
       s_order_cnt = s_order_cnt + 1, 
       s_remote_cnt = s_remote_cnt + ? 
       WHERE s_w_id = ? AND s_i_id = ?;



\end{lstlisting}
\caption{Original version.}
\end{subfigure}
~
\begin{subfigure}[t]{0.48\textwidth}
\begin{lstlisting}
Input: w_id, d_id, array of customer id (c_ids), 
    item_ids, supplier_w_ids, quantities


-- fetch all customer data
1 SELECT c_id, c_discount,... FROM customer
    WHERE c_w_id = ? AND c_d_id = ? AND c_id IN (?,...);
-- fetch warehouse data for all customers
2 SELECT w_tax FROM warehouse WHERE w_id = ?;
-- fetch district d_next_o_id (@next_id) for update
3 SELECT d_next_o_id AS next_id, d_tax FROM district 
    WHERE d_w_id = ? AND d_id = ? FOR UPDATE;
-- increment d_next_o_id by the number of customers (@n)
4 UPDATE district SET d_next_o_id = d_next_o_id + @n
    WHERE d_w_id = ? AND d_id = ?;
-- insert all order and new_order records
5 INSERT INTO oorder (o_id, o_w_id, o_d_id,...) 
    VALUES (@next_id,?,?,...),(@next_id+1,?,?,...),...;
6 INSERT INTO new_order (no_o_id, no_w_id, no_d_id) 
    VALUES (@next_id,?,?),(@next_id+1,?,?),...;

-- execute for all order_line items once
-- fetch all item data
7 SELECT i_id, i_price, i_name, i_data FROM item 
    WHERE i_id IN (?,...);
-- fetch all stock data
8 SELECT s_w_id, s_i_id, s_quantity,... FROM stock 
    WHERE (s_w_id, s_i_id) IN ((?,?),...) FOR UPDATE;
-- insert all order_line records
9 INSERT INTO order_line (ol_w_id, ol_d_id, ol_o_id,...) 
    VALUES (?,?,?,...),...;
-- update all stock
10 UPDATE stock SET 
     s_quantity = CASE WHEN s_w_id = ? AND s_i_id = ? 
         THEN ? ... ELSE s_quantity END, 
     s_ytd = CASE WHEN s_w_id = ? AND s_i_id = ? 
         THEN s_ytd + ? ... ELSE s_ytd END,
     s_order_cnt = CASE WHEN s_w_id = ? AND s_i_id = ? 
         THEN s_order_cnt + 1 ... ELSE s_order_cnt END,
     s_remote_cnt = CASE WHEN s_w_id = ? AND s_i_id = ? 
         THEN s_remote_cnt + ? ... ELSE s_remote_cnt END
     WHERE (s_w_id, s_i_id) IN ((?,?),...);
\end{lstlisting}   
\caption{Merged version.}
\end{subfigure} 

\caption{TPC-C New-Order Transaction.}
\label{fig:neworder-transaction}
\end{figure*}

We merge \code{new-order} transactions to the same warehouse district. As shown in Figure~\ref{fig:neworder-transaction},
the original version takes one customer as the input and the merged version can take multiple customers as the input.
Our static analyzer creates one group for lines 3-4 and one group for lines 8-10 and our investigation
shows that, by re-writing application logic, we can still perform merging within these two groups.
Next, we explain how we perform the merging and its potential benefit line by line:

\begin{itemize}[leftmargin=*]

\item Line 1 of the original version retrieves discount and credit information for one customer. Line 1 of the merged 
version retrieves such information for multiple customers using the \code{select in} semantics as discussed in Section~\ref{sec:intra}.

\item Lines 2 of the original version retrieve the tax  information for future processing. The merged
version only needs to do this once since the tax information is the same for all customers.

\item Lines 3-4 read the next order ID (next\_o\_id) and then increment it by one. These two lines
obviously conflict so our static analyzer groups them. However, instead of executing these
groups sequentially, we find that we can re-write their logic to create more merging opportunity.
Concretely, line 3 of the merged version reads next\_o\_id only once, but it creates an array
[next\_o\_id, next\_o\_id+1, ..., next\_o\_id+n] (used in lines 5-6) to mimic the replies that it would receive
during the sequential execution; line 4 of the merged version
increments next\_o\_id by the number of customers. Such a re-write creates an execution
equivalent to the serial execution but is more efficient.
In the original version, line 4 is a contention point as all customers need
to update this value. In the merged version, however, line 4 becomes an optimization opportunity as contending updates
can be aggregated before executed by the database, as explained in Section~\ref{sec:inter}.
This is one example of modifications that are quite straightforward at the application side but are probably hard to be automated at the database side.

\item Lines 5-6 of the original version insert one open order (oorder) and one new\_order for the customer. Lines 5-6 of the merged version insert multiple orders, one for each customer, using the merged \code{insert} statement.   
Note that these two lines use the mimic next\_o\_id array created in lines 3-4.

\item Lines 7-10 of the original version execute a loop to insert an order\_line for each item and update the item count correspondingly.
Lines 7-10 of the merged version use the merging semantics to replace the loop: Lines 7-8 use \code{select in} semantics to retrieve
information for multiple items; line 9 inserts multiple items into order\_line with the merged \code{insert} statement; line 10 updates the 
information of multiple items using the \code{update case when} statement. Our static analyzer reports a group with lines 8-10
since line 8 and line 10 both access stock quantity. However, we find re-ordering is fine: In the original version, the logic is that
each transaction reads the current stock quantity and increments it by a certain amount; in the merged version, the logic is
that the merged transaction reads the currrent stock quantity and increments by the aggregated amount. These two versions
create an equivalent execution. Note that this is another contention point in the original version.

\end{itemize}

In the merged version, lines 1-6 are effective for inter-transaction merging, and lines 7-10 are effective for both inter- and intra-transaction
merging.

\noindent
\textbf{\code{payment}} transaction processes a payment for an order. 
It identifies the specific warehouse and district where the payment is processed and the customer information who is making the payment. 
Then, it updates the sales statistics associated with the relevant warehouse and district to account for the payment at.
In addition, it updates the customer related fields to reflect the payment.
Finally, it inserts a history record of the payment transaction into the history table.

\begin{figure*}[!t]
\begin{subfigure}[t]{0.48\textwidth}
\begin{lstlisting}
Input: w_id, d_id, c_id, 
    customer payment amount (payment_amount), 
    customer warehouse id (c_w_id), 
    customer district id (c_d_id) 
-- update warehouse w_ytd with payment amount
1 UPDATE warehouse SET w_ytd = w_ytd + ? 
    WHERE w_id = ?;
-- fetch warehouse data
2 SELECT w_street_1,... FROM warehouse WHERE w_id = ?;
-- update district d_ytd with payment amount
3 UPDATE district SET d_ytd = d_ytd + ? 
    WHERE d_w_id = ? AND d_id = ?;
-- fetch district data
4 SELECT d_street_1,... FROM district 
    WHERE d_w_id = ? AND d_id = ?;
-- for simplicity, we only show fetch customer by id
5 SELECT c_credit,... FROM customer 
    WHERE c_w_id = ? AND c_d_id = ? AND c_id = ?;




-- for bad credit customer, fetch c_data and update
if c_credit == "BC":
  -- fetch customer c_data and update
  6 SELECT c_data FROM customer 
      WHERE c_w_id = ? AND c_d_id = ? AND c_id = ?;
  7 UPDATE customer SET 
      c_balance = ?, 
      c_data = ?,... 
      WHERE c_w_id = ? AND c_d_id = ? AND c_id = ?;


-- for good credit customer, update customer balance
else:
  8 UPDATE customer SET 
      c_balance = ?,... 
      WHERE c_w_id = ? AND c_d_id = ? AND c_id = ?;

-- insert history record
9 INSERT INTO history (...) VALUES (...);
\end{lstlisting}
\caption{Original version.}
\end{subfigure}
~
\begin{subfigure}[t]{0.48\textwidth}
\begin{lstlisting}
Input: w_id, d_id, c_ids, 
    array of payment amount (payment_amounts), 
    array of c_w_id (c_w_ids), 
    array of c_d_id (c_d_ids)
-- update warehouse w_ytd with aggregated total payment
1 UPDATE warehouse SET w_ytd = w_ytd + @total 
    WHERE w_id = ?;
-- fetch warehouse data
2 SELECT w_street_1,... FROM warehouse WHERE w_id = ?;
-- update district d_ytd with aggregated total payment
3 UPDATE district SET d_ytd = d_ytd + @total 
    WHERE d_w_id = ? AND d_id = ?;
-- fetch district data
4 SELECT d_street_1,... FROM district 
    WHERE d_w_id = ? AND d_id = ?;
-- fetch all customer data
5 SELECT c_w_id, c_d_id, c_id,c_credit,... FROM customer 
    WHERE (c_w_id, c_d_id, c_id) IN ((?,?,?),...);
-- separate customers into two groups
for each customer c:
    if c.c_credit == "BC": bc_customers.add(c);
    else: gc_customers.add(c);
if bc_customers is not empty:
  -- fetch c_data for bad credit customers and update
  6 SELECT c_w_id, c_d_id, c_id, c_data FROM customer 
      WHERE (c_w_id, c_d_id, c_id) IN ((?,?,?),...);
  7 UPDATE customer SET 
      c_balance = CASE WHEN c_w_id = ? AND c_d_id = ? 
          AND c_id = ? THEN ? ... ELSE c_balance END, 
      c_data = CASE WHEN c_w_id = ? AND c_d_id = ? 
          AND c_id = ? THEN ? ... ELSE c_data END, ... 
      WHERE (c_w_id, c_d_id, c_id) IN ((?,?,?),...);
if gc_customers is not empty:
  -- update for good credit customers
  8 UPDATE customer SET 
      c_balance = CASE WHEN c_w_id = ? AND c_d_id = ? 
          AND c_id = ? THEN ? ... ELSE c_balance END, ... 
      WHERE (c_w_id, c_d_id, c_id) IN ((?,?,?),...);
-- insert all history records
9 INSERT INTO history (...) VALUES (...),...; 
\end{lstlisting}
\caption{Merged version.}
\end{subfigure}

\caption{TPC-C Payment Transaction.}
\label{fig:payment-transaction}
\end{figure*}

Figure~\ref{fig:payment-transaction} shows how we merge the \code{payment} transactions to the same warehouse district.
Our static analyzer reports lines 6-8 as a group since they involve ``if'' statements and thus cannot be automatically
separated, but again, we find ways to re-write them to create more opportunities for merging.

\begin{itemize}[leftmargin=*]

\item Line 1 and line 3 of the original version increments the warehouse  and district year-to-date  balance by the value in the
order. Line 1  and line 3 of the merged version increments this balance by the total values of all orders processed by the merged version.
As transactions to the same warehouse and district need to update these same values, they are
contention points in the original version, and again, are turned into optimization points in the merged version.

\item Line 2 and line 4 of the original version retrieve the warehouse and district information.  Line 2 and line 4
of the merged version do the same once, as the information will be shared by different orders.

\item Line 5 of the original version retrieves information of the customer. Line 5 of the merged version retrieves
information of multiple customers using the \code{select in} statement.

\item Lines 6-8 of the original version handle two types of customers in different ways. To allow merging, the
merged version first separates customers into two groups, each corresponding to one type (the pseudo code 
before line 6). Then within each group, the merged version can apply merging, with \code{select in} statement
and \code{update case when} statements. This is another example of places
we need to tune application logic to a small extent.

\item Line 9 of the original version inserts a record into the history table. Line 9 of the merged version
inserts multiple records with the merged \code{insert} statement.

\end{itemize}

We apply merging to the \code{delivery} transaction as well, but since the pattern to merge its statements is not much 
different from that of \code{new-order} and \code{payment}, we do not present its details here.
The other two types of transactions, \code{order-status} and \code{stock-level},
 are typically not performance bottlenecks, so we do not apply merging
to them.

\subsection{Spree}
\label{sec:spree}
Spree Commerce~\cite{spree} is an open-source e-commerce platform built with Ruby on Rails.
The platform offers a comprehensive set of features to build an online store, 
involving management for products, orders, customers, as well as cart functionality and more.
This platform is popular in the real world and is being used by thousands of businesses around the world, 
among them brands such as GoDaddy, Bookshop.com, KFC, Mitchells, etc~\cite{spreecommerceWhatSpree}.

We deploy a Spree website on our machine and simulate a customer behavior to check which Spree
transactions are triggered. Following this simulation,
we implement merging for two transactions: \code{new-order}, \code{add-item}, and \code{update-stock},
since these three are probably the most frequently used transactions.
Since \code{new-order} and \code{update-stock} are relatively simple, we only present \code{add-item} here.

\noindent
\textbf{\code{add-item}} transaction adds a new item to an existing order in the Spree application.
It retrieves pricing, product details, tax category data, and available stock; 
then it inserts a new line item record; re-calculates totals for the order; and finally updates the order’s summary fields.
Our static analyzer reports a group involving lines 2 and 7 and one involving lines 10-11. We find 
the first group is caused by a false conflict, as line 2 checks the existence of a row and line 7 updates a column
of the row and thus they do not really conflict. For the second group, we find it is fine to re-order its statements.

\begin{figure*}[!t]

\begin{subfigure}[b]{0.48\textwidth}
\begin{lstlisting}
Input: order_id, variant_id, currency, 
    tax_category_id, quantity
-- fetch price for the item
1 SELECT * FROM spree_prices WHERE deleted_at IS NULL 
    AND currency = ? AND variant_id = ? LIMIT 1;
-- check for line item existence
2 SELECT * FROM spree_line_items 
    WHERE order_id = ? AND variant_id = ? 
    ORDER BY created_at ASC LIMIT 1;
-- fetch tax category details
3 SELECT * FROM spree_tax_categories 
    WHERE deleted_at IS NULL AND id = ? LIMIT 1;
-- fetch product details
4 SELECT * FROM spree_products 
    WHERE id = ? LIMIT 1;
-- fetch the stock inventory
5 SELECT SUM(count_on_hand) FROM spree_stock_items 
    INNER JOIN spree_stock_locations 
    ON spree_stock_locations.deleted_at IS NULL 
        AND spree_stock_locations.id = 
            spree_stock_items.stock_location_id 
    WHERE spree_stock_items.deleted_at IS NULL 
        AND spree_stock_locations.deleted_at IS NULL 
        AND spree_stock_locations.active = 1 
        AND variant_id = ?;
-- insert the line item record
6 INSERT INTO spree_line_items (variant_id, order_id, 
    quantity, price, created_at, updated_at, currency) 
    VALUES (?,?,?,?,?,?,?);
-- update the line item record
7 UPDATE spree_line_items SET pre_tax_amount = ? 
    WHERE id = ?;
-- fetch total quantity of the order
8 SELECT SUM(quantity) FROM spree_line_items 
    WHERE order_id = ?;
-- fetch total price of the order
9 SELECT SUM(price * quantity) 
    FROM spree_line_items 
    WHERE order_id = ?;
-- update order record
10 UPDATE spree_orders SET 
     item_total = ?, 
     item_count = ?, 
     total = ?, 
     updated_at = ? 
     WHERE id = ?;
-- ensure the updated_at timestamp is refreshed
11 UPDATE spree_orders SET updated_at = ? 
     WHERE id = ?;
\end{lstlisting}
\caption{Original version.}
\end{subfigure}
~
\begin{subfigure}[b]{0.48\textwidth}
\begin{lstlisting}
Input: order_ids, variant_ids, currencies, 
    tax_category_ids, quantities
-- fetch prices for all items
1 SELECT * FROM spree_prices WHERE deleted_at IS NULL 
    AND currency = ? AND variant_id IN (?,...);
-- check for line item existence
2 SELECT * FROM spree_line_items 
    WHERE (order_id, variant_id) IN ((?,?),...);
-- fetch tax category details
3 SELECT * FROM spree_tax_categories 
    WHERE deleted_at IS NULL AND id IN (?,...);
-- fetch product details
4 SELECT * FROM spree_products WHERE id IN (?,...);
-- fetch stock inventory for all items
5 SELECT variant_id, SUM(count_on_hand) FROM 
    spree_stock_items INNER JOIN spree_stock_locations 
    ON spree_stock_locations.deleted_at IS NULL 
        AND spree_stock_locations.id = 
            spree_stock_items.stock_location_id 
    WHERE spree_stock_items.deleted_at IS NULL 
        AND spree_stock_locations.deleted_at IS NULL 
        AND spree_stock_locations.active = 1 
        AND variant_id IN (?,...) 
    GROUP BY variant_id;
-- insert all line item records
6 INSERT INTO spree_line_items (variant_id, order_id, 
    quantity, price, created_at, updated_at, currency) 
    VALUES ((?,?,?,?,?,?,?,),...);
-- update all line item records
7 UPDATE spree_line_items SET pre_tax_amount = 
    CASE WHEN id = ? THEN ? ... ELSE pre_tax_amount END 
    WHERE id IN (?,...);
-- fetch total quantity of all orders
8 SELECT order_id, SUM(quantity) FROM spree_line_items 
    WHERE order_id IN (?,...) GROUP BY order_id;
-- fetch total price of all orders
9 SELECT order_id, SUM(price * quantity) 
    FROM spree_line_items 
    WHERE order_id IN (?,...) GROUP BY order_id;
-- update all order records
10 UPDATE spree_orders SET 
     item_total = CASE WHEN id = ? THEN ? ... 
        ELSE item_total END, 
     item_count = CASE WHEN id = ? THEN ? ... 
        ELSE item_count END, 
     total = CASE WHEN id = ? THEN ? ... ELSE total END, 
     updated_at = ? WHERE id IN (?,...);
11 UPDATE spree_orders SET updated_at = ? 
     WHERE id IN (?,...);
\end{lstlisting}   
\caption{Merged version.}
\end{subfigure} 

\caption{Spree Add-Item Transaction.}
\label{fig:spree-additem-transaction}
\end{figure*}

\begin{itemize}[leftmargin=*]

\item Lines 1-5 of the original version fetch prices for the item, the line item details if exists, 
tax category details, product details, and item stock inventory of all locations.
Lines 1-5 of the merged version fetch such information for multiple items using the \code{select in} statement.
Note that for line 5, since the statement calculates the sum, it's not easy to split the sum in the merged version.
To address this problem, the merged version applies the \code{group by} clause to let the database separate the results
automatically.

\item Line 6 of the original version inserts a record for one item. Line 6 of the merged version inserts
multiple records using the merged \code{insert} statement.

\item Line 7 of the original version sets the pre-tax value of the item using the \code{update} statement. Line
7 of the merged version sets this value for multiple items using the \code{update case when} statement.

\item Lines 8 and 9 of the original version compute the total number of items and the total price of an
order. Lines 8 and 9 of the merged version compute these values for multiple orders using the \code{select in} statement.
Again, to separate the results automatically, it applies the \code{group by} clause.

\item Lines 10 and 11 of the original version update the order information based on the above computation.
Lines 10 and 11 of the merged version update multiple orders. Note that while line 10 uses \code{update case when},
line 11 can use simple \code{update} since it sets all rows to the same value (i.e., timestamp).
These two lines create a group in our static analysis since they both update the ``update\_at'' column,
but reordering them has no visible effects.

\end{itemize}

Spree is different from TPC-C in a few ways, though they both target online shopping scenarios.
 First, Spree does not have the concept of warehouse or district,
and thus there is no obvious way to partition transactions. Our implementation does not apply any specific
partitioner. Second, Spree does not have much intra-transaction merging opportunity, since its \code{add-item} transaction
only adds a single item.

\subsection{Summary}

As shown in these cases studies, merging can bring performance benefits in three
ways in practice. First, sometimes multiple transactions will retrieve the same
information, and when merged, the merged transaction only needs to retrieve
such information once. Second, sometimes multiple transactions try to update
the same value, causing contention. And by aggregating such updates before
applying them to the database, the merged version turns the contention point
into an optimization point. Finally, merging similar statements with specific
SQL semantics is also possible.

The effort to implement the merged version is admittedly non-trivial, but is manageable
for two reasons. First, we can implement merging incrementally, for a subset of
transactions and for a subset of statements within one transaction. This allows
us to focus on performance-critical transactions and statements. We measure the
effects of such incremental merging on TPC-C in Section~\ref{sec:eval}.
Second, the implementation does not require a high-level re-desgin of the application but requires
mostly local adjustment to individual statements and these local adjustments often
follow a few common patterns. Therefore, we argue that application-side
transaction merging provides a feasible approach for performance improvement,
especially when database-side optimizations are either insufficient or unavailable.

\section{Evaluation}
\label{sec:eval}

We evaluate \sys using TPC-C and Spree workloads as discussed above.
We use one machine to run MySQL server, one machine to run \merger, and
one machine to run clients. We run our experiments on CloudLab~\cite{cloudlab},
using c220g5 machines. Each machine is equipped
with two Intel Xeon Silver 4114 10-core CPUs (20 virtual cores with hyperthreading), 192 GB of RAM, and 
one Intel DC S3500 480 GB 6G SATA SSD. The machines are interconnected 
by a Dual-port Intel X520-DA2 10Gb NIC. We run MySQL server version 8.0.38 and configure it with
8 GB buffer pool size and 4 GB redo log capacity.

\begin{figure*}[t]
    \centering 
    \begin{subfigure}[t]{0.3\textwidth}
    \includegraphics[width=\columnwidth]{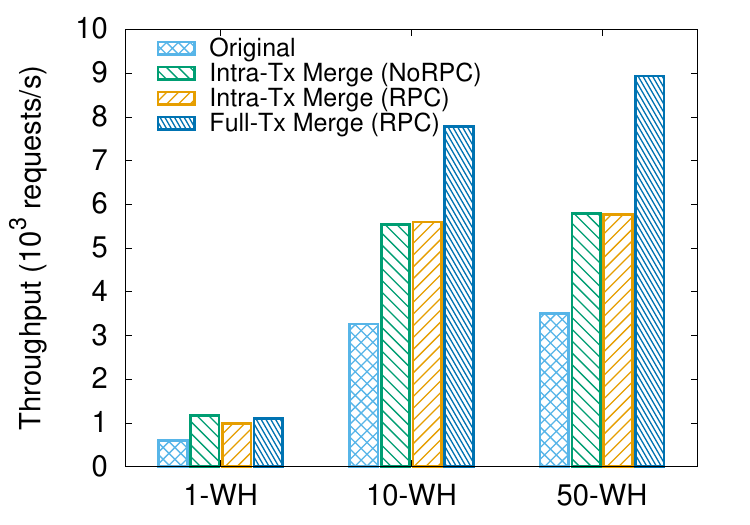}
    \caption{Throughput of TPC-C NewOrder transaction.}
    \label{fig:tpcc-neworder}
    \end{subfigure}
    \hspace{.1in}
    \begin{subfigure}[t]{0.3\textwidth}
    \includegraphics[width=\columnwidth]{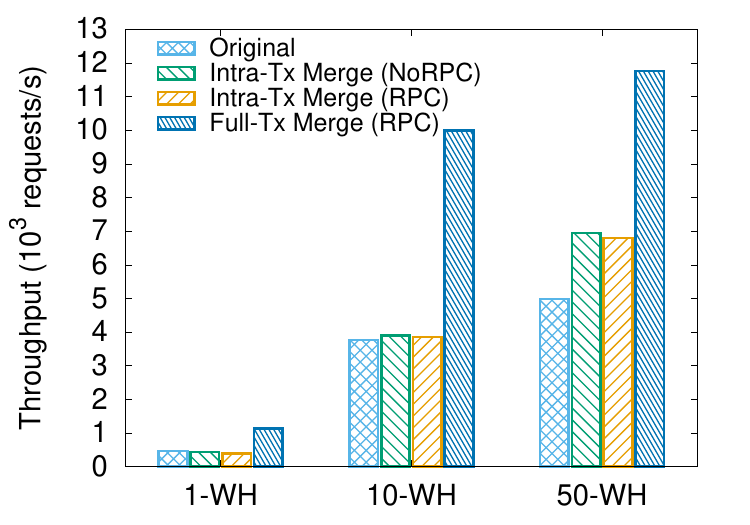}
    \caption{Throughput of TPC-C five transactions.}
    \label{fig:tpcc-alltxn}
    \end{subfigure}
    \hspace{.1in}
    \begin{subfigure}[t]{0.3\textwidth}
    \includegraphics[width=\columnwidth]{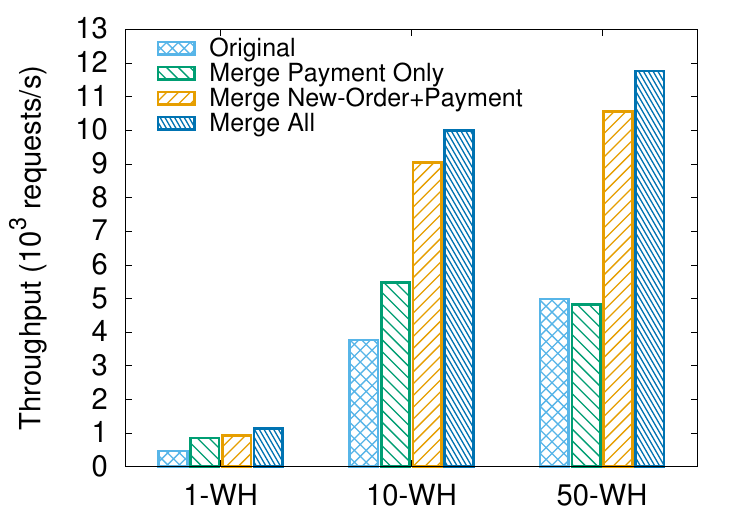}
    \caption{Throughput of TPC-C merging a subset of transactions.}
    \label{fig:tpcc-partial}
    \end{subfigure}
\caption{Throughput of TPC-C on different settings. }
%There are four implementations for the workload transactions.
%The Original, Intra-Tx Merge (NoRPC) refers to the version with intra-transaction merging without RPC, 
%Intra-Tx Merge (RPC) refers to the version with intra-transaction merging with RPC, 
%and Full-Tx Merge (RPC) refers to the version with both intra- and inter-transaction merging.}
\end{figure*}

\vspace{.05in}
\noindent
\textbf{TPC-C throughput.}
For TPC-C, we use the standard data and workload generator from Benchbase~\cite{benchbase}.
We tune the number of worker threads and the number of transactions to merge, and report
the maximal throughput here. We run each experiment three times and report the average number.

Figure~\ref{fig:tpcc-neworder} and Figure~\ref{fig:tpcc-alltxn} measure four settings: the original version with JDBC batching enabled (Original),
the version with intra-transaction merging on the client side (Intra-Tx Merge NoRPC), the version with
the \merger to perform only intra-transaction merging (Intra-Tx Merge RPC) to measure the overhead
of \merger; the full \sys version with both intra- and inter-transaction merging
on the \merger (Full-Tx Merge RPC).

If we run only the \code{new-order} transaction (Figure~\ref{fig:tpcc-neworder}), then
intra-transaction merging alone can already provide a significant improvement.
This is because the \code{new-order} transaction has a number of \code{insert} and \code{update}
statements that can be merged to improve performance (see Figure~\ref{fig:neworder-transaction}).
Note that this improvement can be achieved even without \merger.

The further improvement with inter-transaction merging depends on the setting.
With one warehouse, the improvement is not significant. The reason is that 
two \code{new-order} transactions have two potential contention
points: one is to update the per-district next order ID (d\_next\_o\_id) and the other is to
update the per-warehouse item stock. The first contention will happen when two transactions are to the same
district and the second contention will happen when two transactions purchase the
same item. Since TPC-C uses a skewed distribution to
determine items to purchase, the second is likely to happen.

While our approach to partition transactions based on their warehouse
districts can turn the first contention point into an optimization point, it does not
address the second contention point, as transactions from different districts may purchase
the same item. As a result, with only one warehouse,
the bottleneck is the second contention point and thus the improvement is limited.
With more warehouses, however, transactions to different warehouses can run in parallel.

These experiments demonstrate potential further improvement
opportunities. In this particular example, we may partition transactions based on the tuple
of (district, item), so that we can address two contentions points together. In general,
we may need a generic algorithm to analyze contentions and partition
transactions.

The result with all five types of transactions show a different pattern (Figure~\ref{fig:tpcc-alltxn}).
First, intra-transaction merging does not introduce a significant improvement except in the 50-warehouse case. This is because,
in 1-warehouse and 10-warehouse experiments, the primary bottlenck is the contention in the \code{payment} transactions to
update the year-to-date amount per warehouse (w\_ytd), and intra-transaction merging cannot
address contention problems. Note that this contention is more severe
than those in the \code{new-order} transaction, as \code{payment}'s contention point is the per-warehouse w\_ytd value, \code{new-order}'s
contention point is the per-district d\_next\_o\_id value, and each warehouse has 10 districts.
With 50 warehouses, however, contention is less severe, and intra-transaction
merging can gain its improvement.
Second, inter-transaction merging can consistently improve throughput in all settings, for the same reason.
Since contention in \code{payment} is the primary bottleneck and \code{payment} does not have multiple contention
points like \code{new-order}, it can always benefit from inter-transaction merging to address the contention.
Even for the 50-warehouse case where contention is less severe, inter-transaction merging can still reduce
the number of updates the database needs to execute.

In overall, \sys full version can improve throughput by 1.82X to 2.65X on TPC-C.

Figure~\ref{fig:tpcc-partial} measures the effects of merging a subset of transactions
(the full version merges \code{new-order}, \code{payment}, and \code{delivery}).
As shown in the figure, merging \code{payment} only can get most of the benefit for
the 1-warehouse case; merging \code{new-order} only can get most of the benefit for the
50-warehouse case; for the 10-warehouse case, merging each brings improvement. These experiments show
that a developer can apply merging incrementally to an application, starting from performance-critical
transactions.

\vspace{.05in}
\noindent
\textbf{Spree throughput.} 
There is no standard data and workload generator for Spree, so we created one with our best effort.
To be specific, before an experiment starts, we generate 30000 users, an order for each user, 100000 products,
and a stock with all items. Each order contains 1 item from the stock. We also generate price, 
tax and stock location information using random values within a range. During the experiment,
we run \code{new-order} and \code{add-item} transactions as discussed in Section~\ref{sec:spree}
and we generate their argument values using random values within a range.

\begin{figure}[t]
    \centering
    \includegraphics[width=0.35\textwidth]{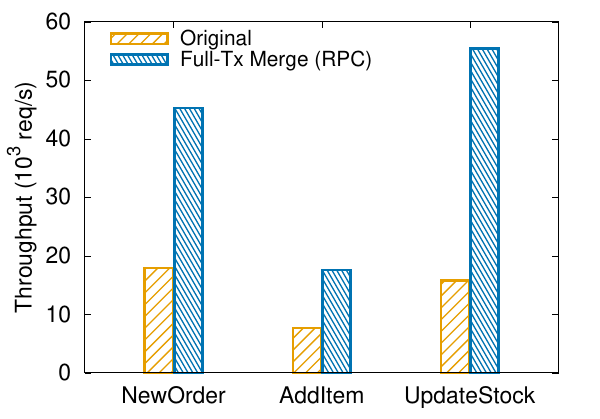}
    \caption{Throughput of Spree.}
    \label{fig:spree-throughput}
\end{figure}

Figure~\ref{fig:spree-throughput} shows the throughput
of Spree. As discussed in Section~\ref{sec:app}, Spree does not have intra-transaction
merging opportunity. Inter-transaction merging still provides a significant improvement,
up to 3.52X.

\begin{figure}[t]
    \centering
    \includegraphics[width=0.45\textwidth]{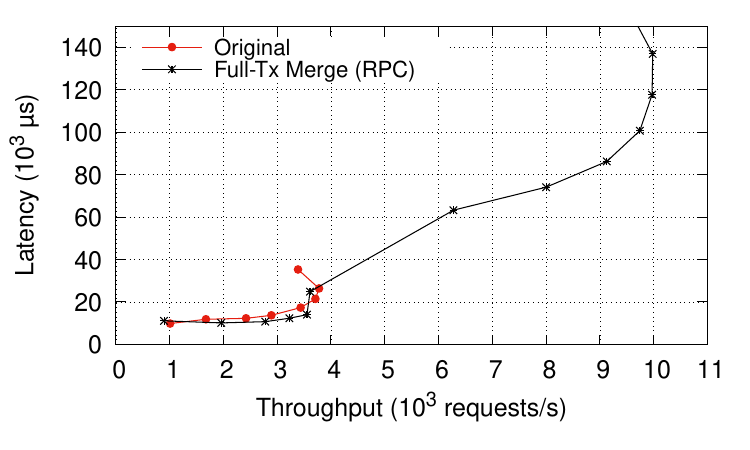}
    \caption{Latency and throughput of TPC-C with 10 warehouses.}
    \label{fig:tpcc-latency-tp}
\end{figure}

\vspace{.05in}
\noindent
\textbf{Impact on latency.}
Since the middleware may add extra latency to transaction processing, 
Figure~\ref{fig:tpcc-latency-tp} measures the latency of processing TPC-C
requests (all five types).  As shown in this figure, the latency of \sys is larger than that of
the original version only when throughput is low (i.e., under 1K requests per second).
When throughput grows higher, \sys's improvement, which helps to reduce queuing
delay, can compensate for the extra latency introduced by the middleware.

Since \sys is designed to improve throughput, it should be applied when throughput
is a concern. When throughput is low, the clients can always bypass \sys and
make direct SQLs to the database (Figure~\ref{fig:design}). And with \sys's support for online re-configuration, 
such settings can be dynamically adjusted at run time.

\vspace{.05in}
\noindent
\textbf{Resource consumption of \merger.}
In \sys, the \merger consumes additional resources (CPU being the major resource in our experiments),
which brings the question whether \sys's approach is worthwhile. In other words,
can we add such additional resource to the database to gain more throughput, instead of using it for a \merger?

\begin{table}
\centering % Center the table horizontally
\caption{Resource consumption vs throughput improvement in TPC-C. Throughput$\uparrow$ is defined as the maximal throughput with \merger
over maximal throughput with only client-side intra-transaction merging. CPU$\uparrow$ is defined as (\merger CPU utilization + database CPU utilization) with \sys over
database CPU utilization with client-side intra-transaction merging.}
\label{tab:resource}
\renewcommand{\arraystretch}{1.3} % Adjust this value to change row height
  \begin{tabular}{ | c | c | c | }
    \hline
    Setting & Throughput$\uparrow$ & CPU$\uparrow$ \\ \hline
    1-WH  & $\frac{1126}{438} = 2.57$   & $\frac{200\% + 400\%}{300\%} = 2.00$    \\ \hline
    10-WH & $\frac{9996}{3893} = 2.57$ & $\frac{1400\% + 2300\%}{1500\%} = 2.47$ \\ \hline
    50-WH & $\frac{11759}{6952} = 1.69$ & $\frac{1600\% + 3000\%}{2800\%} = 1.64$ \\ \hline
  \end{tabular}

\end{table}

Table~\ref{tab:resource} shows some concrete data points.
As shown in this table, for \sys, the throughput improvement is worth
the additional CPU utilization.
On the other hand, since the original version cannot fully utilize the available CPUs,
adding resource to the database side probably won't improve throughput much. This phenomenon
is particularly obvious when contention is severe (e.g., 300\% CPU utilization in 1-WH case,
compared to 4000\% maximal CPU utilization).
In overall, \sys provides a flexible solution compared to adding resources to the database:
\sys can bring improvement under
low-contention and high-contention settings;
due to the stateless nature of \sys,
we can add or remove resource at any time;
the support for online re-configuration allows \sys
to adapt to the workload.
Scaling the database side, however, often involves re-partitioning or migrating
data, which is much more complicated and costly, and won't work well for the
high-contention cases. In practice, we expect these two methods to be complementary.

\vspace{.05in}
\noindent
\textbf{Adjusting to a dynamic workload.}
To evaluate the effectiveness of \sys's online re-configuration, we test whether the Bayesian optimization
can guide the system to an optimal configuration for TPC-C. We randomly set the number of transactions 
to merge from 1 to 10 and the number of threads from 1 to the number of total districts. 
We start from 20 random samples of configuration and run \sys under each sample for 30 seconds to measure
the throughput.
We use the 20 samples to train the GPR model, 
then use the BO to recommend an optimal configuration and retrain the model. 
We repeat the optimization process until convergence, which is defined as the configuration not changing 
and the expected throughput stabilizing. For 1-WH, 10-WH and 50-WH, the convergence 
takes 1, 4, and 2 iterations of optimization after the GPR is trained, respectively.
Of course, a developer may apply her knowledge to narrow down the search space
to accelerate this procedure.

\section{Related Work}
\label{sec:related}

\vspace{.05in}
\noindent
\textbf{Application-side optimization.}
A developer can optimize the performance of a database application
in multiple ways. For example, OLTPShare~\cite{rehrmann2018oltpshare} batches single-statement
read-only transactions to improve throughput in high-load scenarios.
Its follow-up work extends to multi-satement transactions with updates,
but still merges only read operations and executes write normally~\cite{rehrmann2020sharing}.
The benefit of focusing on read-only statements is that these techniques
can be applied automatically at the middle layer, without requiring changes
to the application code. 
Marroquín et. al further applies this idea to queries in cloud-based query-as-a-service (QaaS) systems~\cite{marroquin2018pay}.
It adapts the relational operators to support batched execution, 
and expresses shared execution plan via SQL rewriting, 
using techniques like array operations and unnest operator in SQL constructs.
It can automate the rewriting in simple cases but requires developers' effort
for complex queries.
Our work further explores the benefit of merging
write operations, but requires changes to the application code.

For a contending workload, a common optimization is to split the contending
item into multiple partitions. For example, if a hot item is on sale, quering and
updating its item count may become a contention bottleneck. We can split
the item into several partitions, each holding a portion of the item count, and
let customers go to a random partition. However, load balancing could become
a issue since one partition may be sold out while others are still available.
Our approach avoids the load balancing problem by still maintaining a single
count but merging contending statements before they hit the database.

Other optimization techniques include but are not limited to rewriting a transaction
to move contending statements to later positions when possible,
splitting a transaction into multiple ones, etc~\cite{Shasha1995Chopping,Xie2015HAV}.
These techniques are largely orthogonal to our work.

\vspace{.05in}
\noindent
\textbf{Database-side optimization.} There are a large number of works
optimizing different aspects of a database, such as 
concurrency control~\cite{Thomson2012Calvin, Tu2013Silo, yu2016tictoc, Mu2016Consolidating, ding2018improving, lu2020aria},
commit protocol~\cite{maiyya2019unifying, lu2021epoch}, 
replication~\cite{Thomson2012Calvin, Zhang2015TAPIR, abebe2020dynamast}, 
query planning~\cite{leis2015good, marcus2021bao, zhao2022queryformer}, etc (this is an incomplete
list as there are too many).
The most relevant work we are aware of is SharedDB, which merges multiple queries
into a global plan with shared operators, and processes them together, 
reducing redundant work~\cite{giannikis2012shareddb}.
Our work provides an alternative way to perform such optimizations at the application side.
As discussed, application-side optimizations provide some unique opportunities that
can hardly be automated at the database side.

\vspace{.05in}
\noindent
\textbf{Algorithms to partition transactions.}
To improve database performance, partitioning methods assign transactions to
partitions for concurrent execution, minimize conflicts between partitions
and thus reduce overhead for distributed transactions.
Schism uses a graph-based min-cut partitioning algorithm to partition transactions and balance load, 
minimizes the number of cross-partition edges, i.e. distributed transactions~\cite{curino2010schism}.
Strife groups transactions at runtime into batches, analyzes conflicts via data access graph,
and partitions them into conflict-free clusters and residuals~\cite{prasaad2020handling}.
Its partitioning algorithm operates in three steps: identify k hot data items (one per cluster),
partition transactions into clusters or residuals, merge clusters to reduce residuals.
Horticulture splits tables by frequently accessed columns via access graph,
and assigns transactions to partitions using attribute parameters~\cite{pavlo2012skew}.

Our work proposes to merge contending transactions, when possible. In our experience
with TPC-C, we find simple heuristic-based partitioning is enough in most cases.
With more complex contentions, we anticipate that these algorithms 
can help \sys to determine merging strategy in a generic manner.

\section{Conclusion}

This paper shows that, by merging similar statements
within and across transactions, we can improve
the performance of database applications significantly without
requiring changes to the database backend. 
The required effort is non-trivial but can be applied incrementally.
As a result, transaction merging provides an alternative
for improving database performance, especially when database-side
optimizations are either inefficient or unavailable.

\bibliographystyle{IEEEtran}
\bibliography{LasrBibtex}

\begin{thebibliography}{10}
\providecommand{\url}[1]{#1}
\csname url@samestyle\endcsname
\providecommand{\newblock}{\relax}
\providecommand{\bibinfo}[2]{#2}
\providecommand{\BIBentrySTDinterwordspacing}{\spaceskip=0pt\relax}
\providecommand{\BIBentryALTinterwordstretchfactor}{4}
\providecommand{\BIBentryALTinterwordspacing}{\spaceskip=\fontdimen2\font plus
\BIBentryALTinterwordstretchfactor\fontdimen3\font minus
  \fontdimen4\font\relax}
\providecommand{\BIBforeignlanguage}[2]{{%
\expandafter\ifx\csname l@#1\endcsname\relax
\typeout{** WARNING: IEEEtran.bst: No hyphenation pattern has been}%
\typeout{** loaded for the language `#1'. Using the pattern for}%
\typeout{** the default language instead.}%
\else
\language=\csname l@#1\endcsname
\fi
#2}}
\providecommand{\BIBdecl}{\relax}
\BIBdecl

\bibitem{Thomson2012Calvin}
\BIBentryALTinterwordspacing
A.~Thomson, T.~Diamond, S.-C. Weng, K.~Ren, P.~Shao, and D.~J. Abadi,
  ``{Calvin: Fast Distributed Transactions for Partitioned Database Systems},''
  in \emph{Proceedings of the 2012 ACM SIGMOD International Conference on
  Management of Data}, ser. SIGMOD '12.\hskip 1em plus 0.5em minus 0.4em\relax
  New York, NY, USA: ACM, 2012, pp. 1--12. [Online]. Available:
  \url{http://doi.acm.org/10.1145/2213836.2213838}
\BIBentrySTDinterwordspacing

\bibitem{Tu2013Silo}
\BIBentryALTinterwordspacing
S.~Tu, W.~Zheng, E.~Kohler, B.~Liskov, and S.~Madden, ``{Speedy Transactions in
  Multicore In-memory Databases},'' in \emph{Proceedings of the Twenty-Fourth
  ACM Symposium on Operating Systems Principles}, ser. SOSP '13.\hskip 1em plus
  0.5em minus 0.4em\relax New York, NY, USA: ACM, 2013, pp. 18--32. [Online].
  Available: \url{http://doi.acm.org/10.1145/2517349.2522713}
\BIBentrySTDinterwordspacing

\bibitem{leis2015good}
V.~Leis, A.~Gubichev, A.~Mirchev, P.~Boncz, A.~Kemper, and T.~Neumann, ``How
  good are query optimizers, really?'' \emph{Proceedings of the VLDB
  Endowment}, vol.~9, no.~3, pp. 204--215, 2015.

\bibitem{Zhang2015TAPIR}
I.~Zhang, N.~K. Sharma, A.~Szekeres, A.~Krishnamurthy, and D.~R.~K. Ports,
  ``Building consistent transactions with inconsistent replication,'' ser. SOSP
  '15.\hskip 1em plus 0.5em minus 0.4em\relax New York, NY, USA: Association
  for Computing Machinery, 2015, p. 263–278.

\bibitem{Shasha1995Chopping}
D.~Shasha, F.~Llirbat, E.~Simon, and P.~Valduriez, ``Transaction chopping:
  algorithms and performance studies,'' \emph{ToDS}, 1995.

\bibitem{Xie2015HAV}
\BIBentryALTinterwordspacing
C.~Xie, C.~Su, C.~Littley, L.~Alvisi, M.~Kapritsos, and Y.~Wang,
  ``{High-performance ACID via Modular Concurrency Control},'' in
  \emph{Proceedings of the 25th Symposium on Operating Systems Principles},
  ser. SOSP '15.\hskip 1em plus 0.5em minus 0.4em\relax New York, NY, USA: ACM,
  2015, pp. 279--294. [Online]. Available:
  \url{http://doi.acm.org/10.1145/2815400.2815430}
\BIBentrySTDinterwordspacing

\bibitem{rehrmann2020sharing}
R.~Rehrmann, C.~Binnig, A.~B\"{o}hm, K.~Kim, and W.~Lehner, ``{Sharing
  opportunities for OLTP workloads in different isolation levels},''
  \emph{Proc. VLDB Endow.}, vol.~13, no.~10, p. 1696–1708, Jun. 2020.

\bibitem{dbcc-cidr-2023}
C.~Cheng, M.~Han, N.~Xu, S.~Blanas, M.~D. Bond, and Y.~Wang, ``Developer's
  responsibility or database's responsibility? rethinking concurrency control
  in databases,'' in \emph{CIDR}, 2023.

\bibitem{tang2022adhoc}
C.~Tang, Z.~Wang, X.~Zhang, Q.~Yu, B.~Zang, H.~Guan, and H.~Chen, ``Ad hoc
  transactions in web applications: The good, the bad, and the ugly,'' in
  \emph{SIGMOD}, 2022.

\bibitem{oracleStatementJava}
``Statement (java platform se 8)-executebatch,''
  \url{https://docs.oracle.com/javase/8/docs/api/java/sql/Statement.html}.

\bibitem{tpccurl}
\BIBentryALTinterwordspacing
{Transaction Processing Performance Council}, ``The {TPC-C} home page,''
  http://www.tpc.org/tpcc/. [Online]. Available:
  \url{\url{http://www.tpc.org/tpcc/}}
\BIBentrySTDinterwordspacing

\bibitem{rehrmann2023merging}
R.~Rehrmann, ``{Merging Queries in OLTP Workloads},'' 2023.

\bibitem{spreecommerceWhatSpree}
``{Spree Commerce Documentation},''
  \url{https://spreecommerce.org/docs/user/what-is-spree-commerce}, 2024.

\bibitem{alipourfard2017cherrypick}
O.~Alipourfard, H.~H. Liu, J.~Chen, S.~Venkataraman, M.~Yu, and M.~Zhang,
  ``{CherryPick}: Adaptively unearthing the best cloud configurations for big
  data analytics,'' in \emph{14th USENIX Symposium on Networked Systems Design
  and Implementation (NSDI 17)}, Boston, MA, Mar. 2017, pp. 469--482.

\bibitem{behzadOptimizingPerformanceHPC2018}
\BIBentryALTinterwordspacing
B.~Behzad, S.~Byna, Prabhat, and M.~Snir, ``Optimizing {{I}}/{{O Performance}}
  of {{HPC Applications}} with {{Autotuning}},'' \emph{ACM Trans. Parallel
  Comput.}, vol.~5, no.~4, Mar. 2019. [Online]. Available:
  \url{https://doi.org/10.1145/3309205}
\BIBentrySTDinterwordspacing

\bibitem{agarwalActiveLearningbasedAutomatic2019}
M.~Agarwal, D.~Singhvi, P.~Malakar, and S.~Byna, ``Active {{Learning-based
  Automatic Tuning}} and {{Prediction}} of {{Parallel I}}/{{O Performance}},''
  in \emph{2019 {{IEEE}}/{{ACM Fourth International Parallel Data Systems
  Workshop}} ({{PDSW}})}, Nov. 2019, pp. 20--29.

\bibitem{rajeshTunIOAIpoweredFramework2024}
N.~Rajesh, K.~Bateman, J.~L. Bez, S.~Byna, A.~Kougkas, and X.-H. Sun,
  ``{{TunIO}}: {{An AI-powered Framework}} for {{Optimizing HPC I}}/{{O}},'' in
  \emph{2024 {{IEEE International Parallel}} and {{Distributed Processing
  Symposium}} ({{IPDPS}})}.\hskip 1em plus 0.5em minus 0.4em\relax IEEE, pp.
  494--505.

\bibitem{grpcGRPC}
``g{R}{P}{C},'' \url{https://grpc.io/}.

\bibitem{GPR}
``{G}aussian{P}rocess{R}egressor with scikit-learn,''
  \url{https://scikit-learn.org/stable/modules/generated/sklearn.gaussian\_process.GaussianProcessRegressor.html}.

\bibitem{BO}
``{Bayesian Optimization with skopt},''
  \url{https://scikit-optimize.github.io/stable/auto_examples/bayesian-optimization.html}.

\bibitem{benchbase}
``Benchbase,'' \url{https://github.com/cmu-db/benchbase}.

\bibitem{spree}
``{Spree},'' \url{https://github.com/spree/spree}.

\bibitem{cloudlab}
``{{CloudLab}},'' \url{https://www.cloudlab.us}.

\bibitem{rehrmann2018oltpshare}
R.~Rehrmann, C.~Binnig, A.~B\"{o}hm, K.~Kim, W.~Lehner, and A.~Rizk,
  ``{OLTPshare: the case for sharing in OLTP workloads},'' \emph{Proc. VLDB
  Endow.}, vol.~11, no.~12, p. 1769–1780, Aug. 2018.

\bibitem{marroquin2018pay}
R.~Marroqu\'{\i}n, I.~M\"{u}ller, D.~Makreshanski, and G.~Alonso, ``Pay one,
  get hundreds for free: Reducing cloud costs through shared query execution,''
  in \emph{Proceedings of the ACM Symposium on Cloud Computing}, ser. SoCC
  '18.\hskip 1em plus 0.5em minus 0.4em\relax New York, NY, USA: Association
  for Computing Machinery, 2018, p. 439–450.

\bibitem{yu2016tictoc}
X.~Yu, A.~Pavlo, D.~Sanchez, and S.~Devadas, ``{TicToc: Time Traveling
  Optimistic Concurrency Control},'' in \emph{Proceedings of the 2016
  International Conference on Management of Data}, ser. SIGMOD '16.\hskip 1em
  plus 0.5em minus 0.4em\relax New York, NY, USA: Association for Computing
  Machinery, 2016, p. 1629–1642.

\bibitem{Mu2016Consolidating}
\BIBentryALTinterwordspacing
S.~Mu, L.~Nelson, W.~Lloyd, and J.~Li, ``{Consolidating Concurrency Control and
  Consensus for Commits under Conflicts},'' in \emph{12th {USENIX} Symposium on
  Operating Systems Design and Implementation ({OSDI} 16)}.\hskip 1em plus
  0.5em minus 0.4em\relax Savannah, GA: {USENIX} Association, 2016, pp.
  517--532. [Online]. Available:
  \url{https://www.usenix.org/conference/osdi16/technical-sessions/presentation/mu}
\BIBentrySTDinterwordspacing

\bibitem{ding2018improving}
B.~Ding, L.~Kot, and J.~Gehrke, ``Improving optimistic concurrency control
  through transaction batching and operation reordering,'' \emph{Proceedings of
  the VLDB Endowment}, vol.~12, no.~2, pp. 169--182, 2018.

\bibitem{lu2020aria}
Y.~Lu, X.~Yu, L.~Cao, and S.~Madden, ``{Aria: a fast and practical
  deterministic OLTP database},'' \emph{Proc. VLDB Endow.}, vol.~13, no.~12, p.
  2047–2060, 2020.

\bibitem{maiyya2019unifying}
S.~Maiyya, F.~Nawab, D.~Agrawal, and A.~E. Abbadi, ``Unifying consensus and
  atomic commitment for effective cloud data management,'' \emph{Proceedings of
  the VLDB Endowment}, vol.~12, no.~5, pp. 611--623, 2019.

\bibitem{lu2021epoch}
Y.~Lu, X.~Yu, L.~Cao, and S.~Madden, ``{Epoch-based commit and replication in
  distributed OLTP databases},'' 2021.

\bibitem{abebe2020dynamast}
M.~Abebe, B.~Glasbergen, and K.~Daudjee, ``Dynamast: Adaptive dynamic mastering
  for replicated systems,'' in \emph{2020 IEEE 36th international conference on
  data engineering (ICDE)}.\hskip 1em plus 0.5em minus 0.4em\relax IEEE, 2020,
  pp. 1381--1392.

\bibitem{marcus2021bao}
R.~Marcus, P.~Negi, H.~Mao, N.~Tatbul, M.~Alizadeh, and T.~Kraska, ``Bao:
  Making learned query optimization practical,'' in \emph{Proceedings of the
  2021 International Conference on Management of Data}, 2021, pp. 1275--1288.

\bibitem{zhao2022queryformer}
Y.~Zhao, G.~Cong, J.~Shi, and C.~Miao, ``Queryformer: A tree transformer model
  for query plan representation,'' \emph{Proceedings of the VLDB Endowment},
  vol.~15, no.~8, pp. 1658--1670, 2022.

\bibitem{giannikis2012shareddb}
G.~Giannikis, G.~Alonso, and D.~Kossmann, ``{SharedDB: killing one thousand
  queries with one stone},'' \emph{Proc. VLDB Endow.}, vol.~5, no.~6, p.
  526–537, Feb. 2012.

\bibitem{curino2010schism}
C.~Curino, E.~Jones, Y.~Zhang, and S.~Madden, ``Schism: a workload-driven
  approach to database replication and partitioning,'' \emph{Proc. VLDB
  Endow.}, vol.~3, no. 1–2, p. 48–57, Sep. 2010.

\bibitem{prasaad2020handling}
G.~Prasaad, A.~Cheung, and D.~Suciu, ``{Handling highly contended OLTP
  workloads using fast dynamic partitioning},'' in \emph{Proceedings of the
  2020 ACM SIGMOD International Conference on Management of Data}, 2020, pp.
  527--542.

\bibitem{pavlo2012skew}
A.~Pavlo, C.~Curino, and S.~Zdonik, ``{Skew-aware automatic database
  partitioning in shared-nothing, parallel OLTP systems},'' in
  \emph{Proceedings of the 2012 ACM SIGMOD International Conference on
  Management of Data}, 2012, pp. 61--72.

\end{thebibliography}

\end{document}